\documentclass[runningheads,a4paper]{llncs}
\usepackage{amssymb}
\usepackage{graphicx}

\usepackage{url}
\urldef{\mailsa}\path|r.k.moore@sheffield.ac.uk|
\newcommand{\keywords}[1]{\par\addvspace\baselineskip
\noindent\keywordname\enspace\ignorespaces#1}

\begin{document}

\mainmatter  

\title{Is Spoken Language All-or-Nothing?  Implications for future speech-based human-machine interaction}

\titlerunning{Is Spoken Language All-or-Nothing?}

%
%
\author{Roger K. Moore}
\authorrunning{Is Spoken Language All-or-Nothing?}

\institute{Speech and Hearing Research Group, \\Dept. Computer Science, University of Sheffield, \\Regent Court, 211 Portobello, Sheffield, S1 4DP, UK\\
\mailsa\\
\url{http://www.dcs.shef.ac.uk/~roger/}}

%
%

\toctitle{Is Spoken Language All-or-Nothing?}
\tocauthor{Roger K. Moore}
\maketitle

\begin{abstract}

Recent years have seen significant market penetration for voice-based personal assistants such as Apple's \emph{Siri}.  However, despite this success, user take-up is frustratingly low.  This position paper argues that there is a \emph{habitability gap} caused by the inevitable mismatch between the capabilities and expectations of human users and the features and benefits provided by contemporary technology.  Suggestions are made as to how such problems might be mitigated, but a more worrisome question emerges: ``\emph{is spoken language all-or-nothing}''?  The answer, based on contemporary views on the special nature of (spoken) language, is that there may indeed be a fundamental limit to the interaction that can take place between mismatched interlocutors (such as humans and machines).  However, it is concluded that interactions between native and non-native speakers, or between adults and children, or even between humans and dogs, might provide critical inspiration for the design of future speech-based human-machine interaction.


\end{abstract}

\section{Introduction} \label{sec:INT}

The release in 2011 of \emph{Siri}, Apple's voice-based personal assistant for the iPhone, signalled a step change in the public perception of spoken language technology.  For the first time, a significant number of everyday users were exposed to the possibility of using their voice to enter information, navigate applications or pose questions - all by speaking to their mobile device.  Of course, voice dictation software had been publicly available since the release of \emph{Dragon Naturally Speaking} in 1997, but such technology only found success in niche market areas for document creation (by users who could not or would not type).  In contrast, \emph{Siri} appeared to offer a more general-purpose interface that thrust the potential benefits of automated speech-based interaction into the forefront of the public's imagination.  By combining automatic speech recognition and speech synthesis with natural language processing and dialogue management, \emph{Siri} promoted the possibility of a more \emph{conversational} interaction between users and smart devices.  As a result, competitors such as \emph{Google Now} and Microsoft's \emph{Cortana} soon followed\footnote{See \cite{Pieraccini2012} for a comprehensive review of the history of speech technology R\&D up to, and including, the release of \emph{Siri}.}.

Of course, it is well established that, while voice-based personal assistants such as \emph{Siri} are now very familiar to the majority of mobile device users, their practical value is still in doubt.  This is evidenced by the preponderance of videos on \emph{YouTube}\textsuperscript{TM} that depict humorous rather than practical uses; it seems that people give such systems a try, play around with them for a short while and then go back to their more familiar ways of doing things.  Indeed, this has been confirmed by a recent survey of users from around the world which showed that only 13\% of the respondents used a facility such as \emph{Siri} every day, whereas 46\% had tried it once and then given up (citing inaccuracy and a lack of privacy as key reasons for abandoning it) \cite{Liao2015}.

This lack of serious take-up of voice-based personal assistants could be seen as the inevitable teething problems of a new(ish) technology, or it could be evidence of something more deep-seated.  This \emph{position} paper addresses these issues, and attempts to tease out some of the overlooked features of spoken language that might have a bearing on the success or failure of voice-based human-machine interaction.  In particular, attention is drawn to the inevitable \emph{mismatch} between the capabilities and expectations of human users and the features and benefits provided by contemporary technical solutions.  Some suggestions are made as to how such problems might be mitigated, but a more worrisome question emerges: ``\emph{is spoken language all-or-nothing}''?

\section{The Nature of the Problem} \label{sec:PRB}

\begin{figure}
\centering
\includegraphics[width=12cm]{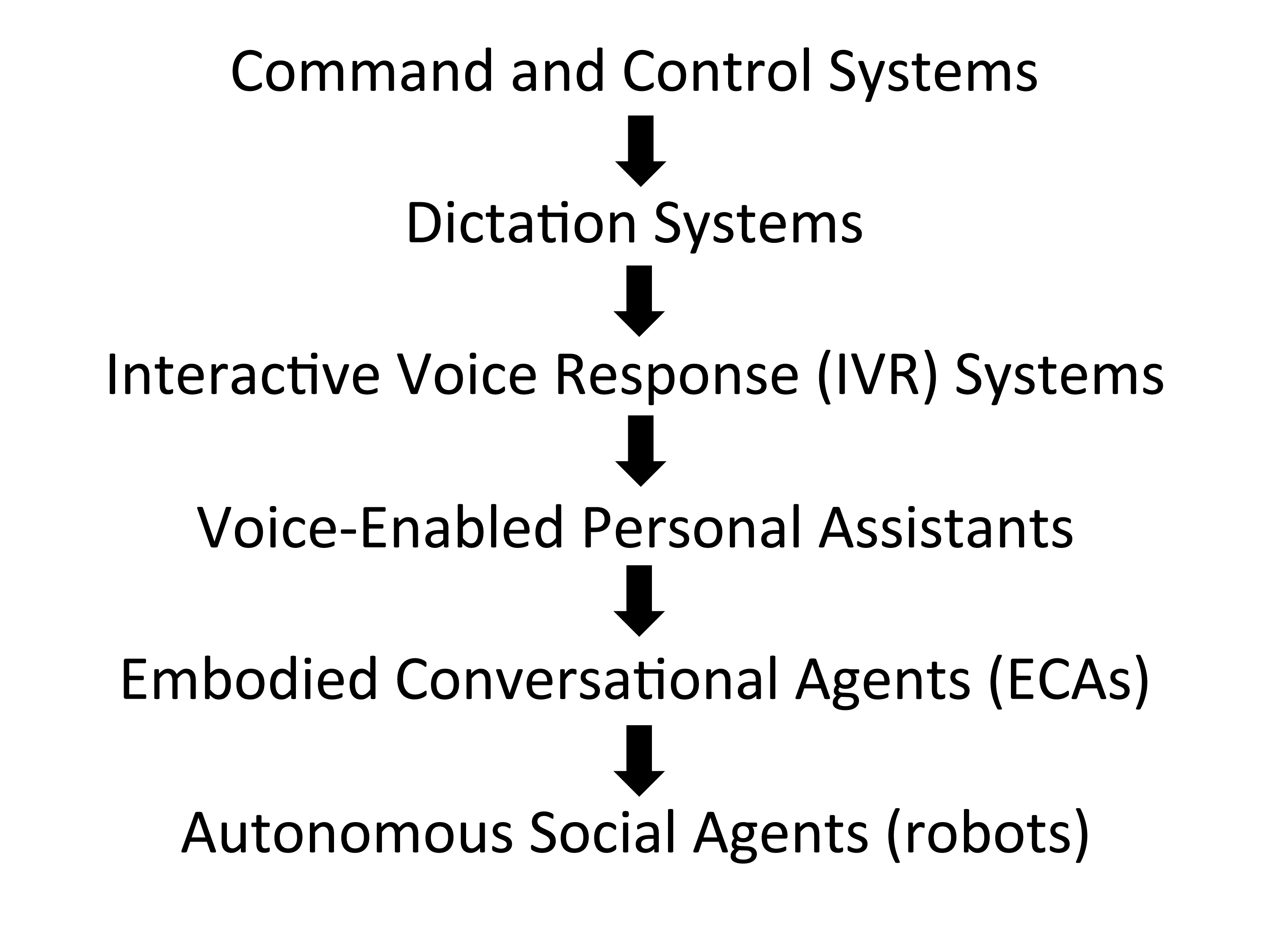}
\caption{The evolution of spoken language technology applications from early military \emph{Command and Control Systems} to future \emph{Autonomous Social Agents} (robots).}
\label{fig:DEV}
\end{figure}

There are many challenges facing the development of effective voice-based human-machine interaction \cite{Deng2004,Minker2007}.  As the technology has matured, so the applications that are able to be supported have grown in depth and complexity (see Fig.\ref{fig:DEV}).  From the earliest military \emph{Command and Control Systems} to contemporary commercial \emph{Interactive Voice Response (IVR) Systems} and the latest \emph{Voice-Enabled Personal Assistants} (such as \emph{Siri}), the variety of human accents, competing signals in the acoustic environment and the complexity of the application scenario have always presented significant barriers to practical usage.  Considerable progress has been made in all of the core technologies, particularly following the emergence of the data-driven stochastic modelling paradigm \cite{Gales2007} (now supplemented by \emph{deep learning} \cite{Hinton2012}) as a key driver in pushing regularly benchmarked performance in a positive direction.  Yet, as we have seen, usage remains a serious issue; not only does a speech interface compete with very effective non-speech GUIs \cite{Moore2004}, but people have a natural aversion to talking to machines in public spaces \cite{Liao2015}.  As Nass \& Brave stated in their seminal book \emph{Wired for Speech} \cite{Nass2005}: ``\emph{voice interfaces have become notorious for fostering frustration and failure}'' (p.6).

These problems become magnified as the field moves forward to developing voice-based interaction with \emph{Embodied Conversational Agents (ECAs)} and \emph{Autonomous Social Agents} (robots).  In these futuristic scenarios, it is assumed that spoken language will provide a ``natural'' conversational interface between human beings and so-called \emph{intelligent} systems.  However, there many additional challenges which need to be overcome in order to address such a requirement \ldots

\begin{quotation}
``\emph{We need to move from developing robots that simply talk and listen to evolving intelligent communicative machines that are capable of truly understanding human behaviour, and this means that we need to look beyond speech, beyond words, beyond meaning, beyond communication, beyond dialogue and beyond one-off interactions.}'' \cite{Moore2015} (p.321)
\end{quotation}

\begin{figure}
\centering
\includegraphics[width=12cm]{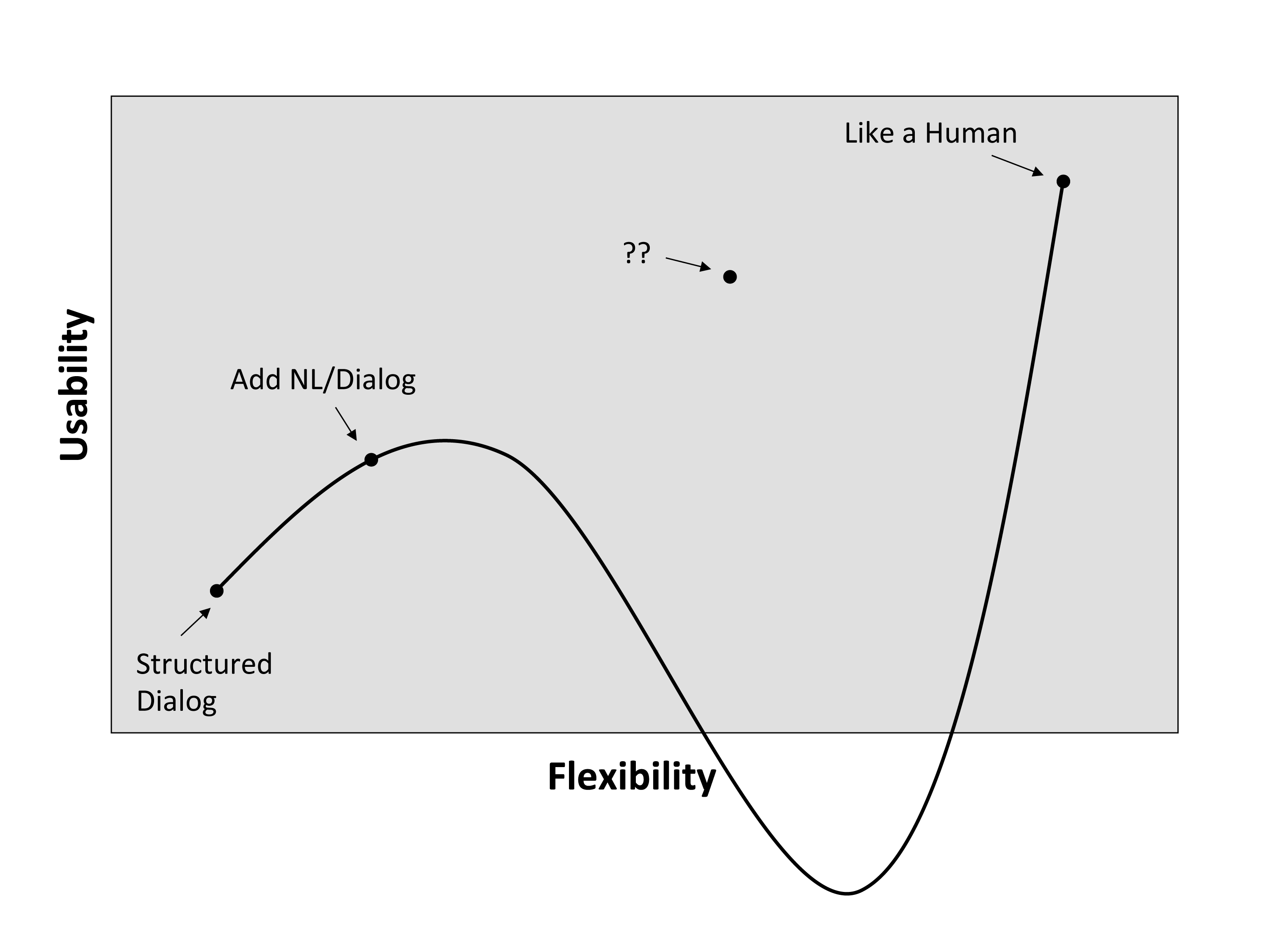}
\caption{Illustration of the consequences of increasing the flexibility of spoken language dialogue systems; increasing flexibility can lead to a \emph{habitability gap} where usability drops catastrophically (reproduced, with permission, from Mike Phillips \cite{Philips2006}).  This means that it is surprisingly difficult to deliver a technology corresponding to the point marked `??'.  \emph{Siri} corresponds to the point marked `Add NL/Dialog'.}
\label{fig:MP}
\end{figure}

Of these, a perennial problem seems to be how to evolve the complexity of voice-based interfaces from simple structured dialogues to more flexible conversational designs without confusing the user \cite{Bernsen1998,McTear2004,Lopez2005}.  Indeed, it has been known for some time that there appears to be a non-linear relationship between \emph{flexibility} and \emph{usability} \cite{Philips2006} - see Fig.\ref{fig:MP}.  As flexibility increases with advancing technology, so usability increases until users no longer know what they can and cannot say, at which point usability tumbles and interaction falls apart.

\subsection{The ``Habitability Gap''} \label{sec:HG}

Progress is being made in this area: for example, by providing targeted help to users \cite{Tomko2005,Tomko2006,Komatani2007} and by replacing the traditional notion of turn-taking with a more fluid interaction based on \emph{incremental processing} \cite{Schlangen2009,Hastie2012}.  Likewise, simple slot-filling approaches to language understanding and generation are being replaced by sophisticated statistical methods for estimating dialogue states and optimal next moves \cite{Williams2007,Gasic2013}.  Nevertheless, it is still the case that there is a \emph{habitability gap} of the form illustrated in Fig.\ref{fig:MP}.

In fact, the shape of the curve illustrated in Fig.\ref{fig:MP} is virtually identical to the famous \emph{Uncanny Valley effect} \cite{Mori1970} in which a near human-looking artefact (such as a humanoid robot) can trigger feelings of eeriness and repulsion in an observer; as \emph{human likeness} increases, so \emph{affinity} increases until a point where artefacts start to appear creepy and affinity goes negative.  A wide variety of explanations have been suggested for this non-linear relationship but, to date, there is only one quantitative model \cite{Moore2012}, and this is founded on the combined effect of categorical perception and mismatched perceptual cues giving rise to a form of \emph{perceptual tension}.  The implication of this model is that uncanniness - and hence, habitability - can be avoided if care is taken to align how an autonomous agent looks, sounds and behaves \cite{MooreMaier2012,Moore2015}.  In other words, if a speech-enabled agent is to converse successfully with a human being, it should make clear its interactional \emph{affordances} \cite{Gibson1977,Worgan2010}.

This analysis leads to an important implication - since a spoken language system consists of a number of different components, each of which possesses a certain level of technical capability, then in order to be coherent (and hence \emph{usable}), the design of the overall system needs to be aligned to the component with the \emph{lowest} level of performance.  For example, giving an automated personal assistant a natural human voice is a recipe for user confusion in the (normal) situation where the other speech technology components are limited in their abilities.  In other words, in order to maximise the effectiveness of the interaction, \textbf{a speech-enabled robot should have a robot voice}.  As Bruce Balentine succinctly puts it \cite{Balentine2007}: ``\textit{It's better to be a good machine than a bad person}''!  This is an unpopular result\footnote{It is often argued that such an approach is unimportant as users will habituate.  However, habituation only occurs after sustained exposure, and a key issue here is how to increase the effectiveness of first encounters (since that has a direct impact on the likelihood of further usage).}, but there is evidence of its effectiveness \cite{Moore1992}, and it clearly has implications for contemporary voice-based personal assistants such as \emph{Siri}, \emph{Google Now} and \emph{Cortana} which employ very humanlike voices\footnote{Interestingly, these ideas do appear to be having some impact on the design of contemporary autonomous social agents such as \emph{Jibo} (which has a childlike and mildly robotic voice) \cite{Jibo2015}.}.

Of course, some might claim that the habitability problem only manifests itself in applications where task-completion is a critical measure of success.  The suggestion would be that the situation might be different for applications in domains such as social robots, education or games in which the emphasis would be more on the spoken interaction itself.  However, the argument presented in this paper is not concerned with the nature of the interaction, rather it questions whether such speech-based interaction can be sustained without access to the notion of \emph{full} language.

\subsection{Half a Language?} \label{sec:HLF}

So far, so good - as component technologies improve, so the flexibility of the overall system would increase, and as long as the capabilities of the individual components are aligned, it should be possible to avoid falling into the habitability gap.

However, sending mixed messages about the capabilities of a spoken language system is only one part of the story; even if a speech-based autonomous social agent looks, sounds and behaves in a coherent way, will users actually be able to engage in conversational interaction if the overall capability is \emph{less} than that normally enjoyed by a human being?  What does it mean for a language-based system to be compromised in some way?  How can users know what they may and may not say \cite{Jokinen2006,Tomko2006}, or even if this is the right question?  Is there such a thing as \emph{half} a language and, if so, is it habitable?  Indeed, what is language anyway?

\section{What is Language?} \label{sec:LNG}

Unfortunately there is no space here to review the extensive and, at times, controversial history of the scientific study of language, or of the richness and variety of its spoken (and gestural) forms.  Suffice to say that human beings have evolved a prolific system of (primary vocal) interactive behaviours that is vastly superior to that enjoyed by any other animal \cite{Gardiner1932,Bickerton1995,Hauser1997,Hauser2002,Everett2012}.   As has been said a number of times \ldots

\begin{quotation}
``\emph{Spoken language is the most sophisticated behaviour of the most complex organism in the known universe.}'' \cite{Moore2007}.
\end{quotation}

The complexity and sophistication of (spoken) language tends to be masked by the apparent ease with which we, as human beings, use it.  As a consequence, engineered solutions are often dominated by a somewhat na{\"i}ve perspective involving the coding and decoding of messages passing from one brain (the sender) to another brain (the receiver).  In reality, \emph{languaging} is better viewed as an emergent property of the dynamic coupling between \emph{cognitive unities} that serves to facilitate distributed sense-making through cooperative behaviours and, thereby, social structure \cite{Maturana1987,Cummins2004,Bickhard2007,Cowley2011,Fusaroli2014}.  Furthermore, the contemporary view is that language is based on the co-evolution of two key traits - \emph{ostensive-inferential} communication and \emph{recursive mind-reading} (including `Theory-of-Mind') \cite{ScottPhillips2015,Baron1999,Malle2002} - and that abstract (mental) meaning is grounded in the concrete (physical) world through \emph{metaphor} \cite{Lakoff1980,Feldman2008}.

These modern perspectives on language not only place strong emphasis on \emph{pragmatics} \cite{Levinson1983}, but they are also founded on an implicit assumption that interlocutors are conspecifics\footnote{Members of the same species.} and hence share significant priors.  Indeed, evidence suggests that some animals draw on representations of their own abilities (expressed as predictive models \cite{Friston2009}) in order to interpret the behaviours of others \cite{Rizzolatti2004,Wilson2005}.  For human beings, this is thought to be a key enabler for efficient recursive mind-reading and hence for language \cite{Pickering2007,Garrod2013}.

\begin{figure}
\centering
\includegraphics[width=12cm]{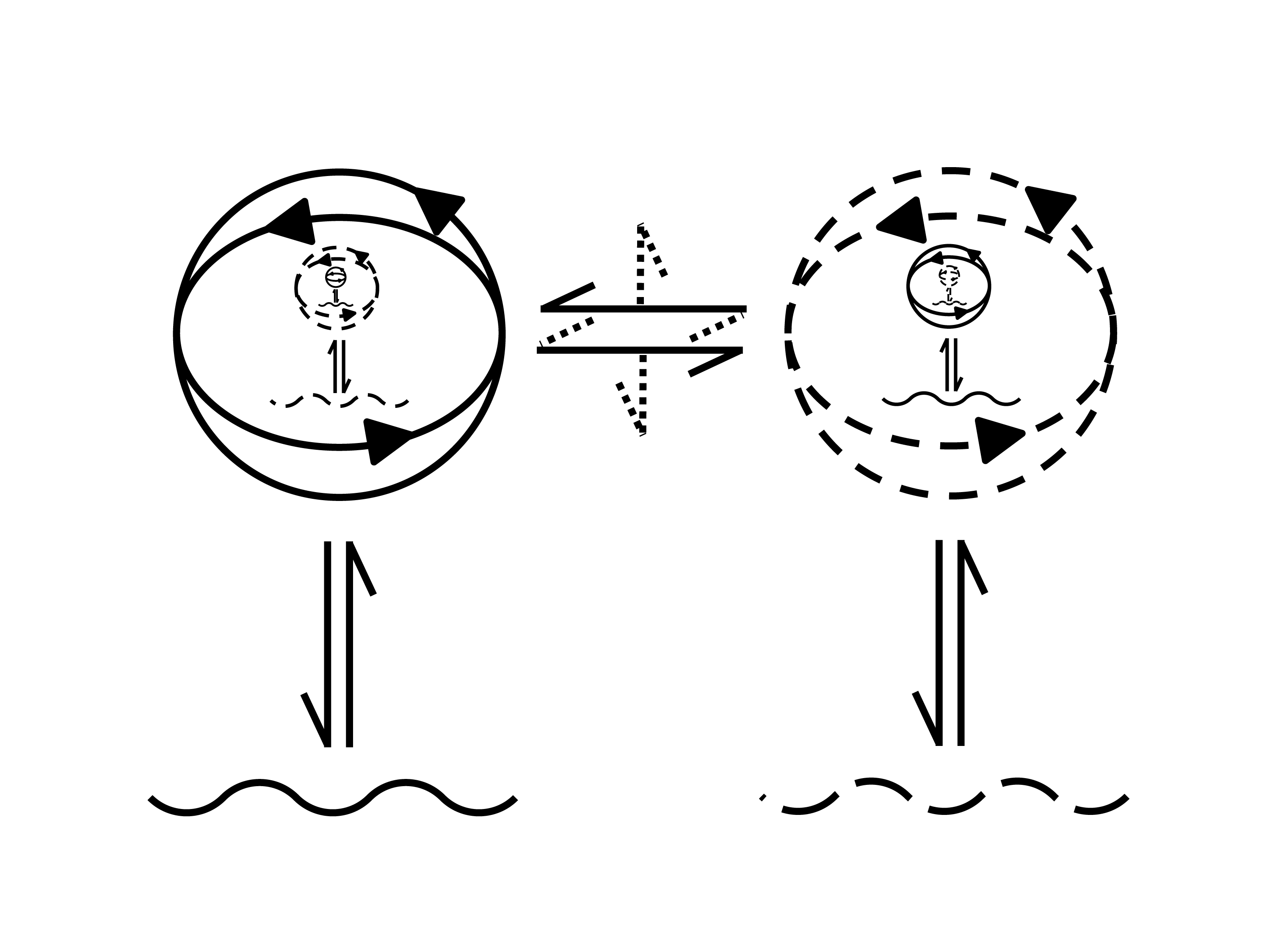}
\caption{Pictographic representation of language-based coupling (dialogue) between two human interlocutors \cite{Moore2016}.  One interlocutor (and its environment) is depicted using solid lines and the other interlocutor (and its environment) is depicted using broken lines.  As can be seen, communicative interaction is founded on two-way ostensive recursive mind-reading (including mutual Theory-of-Mind).}
\label{fig:DIA}
\end{figure}

Several of these advanced concepts may be usefully expressed in pictographic form \cite{Moore2016} - see Fig.\ref{fig:DIA}.

\begin{figure}
\centering
\includegraphics[width=12cm]{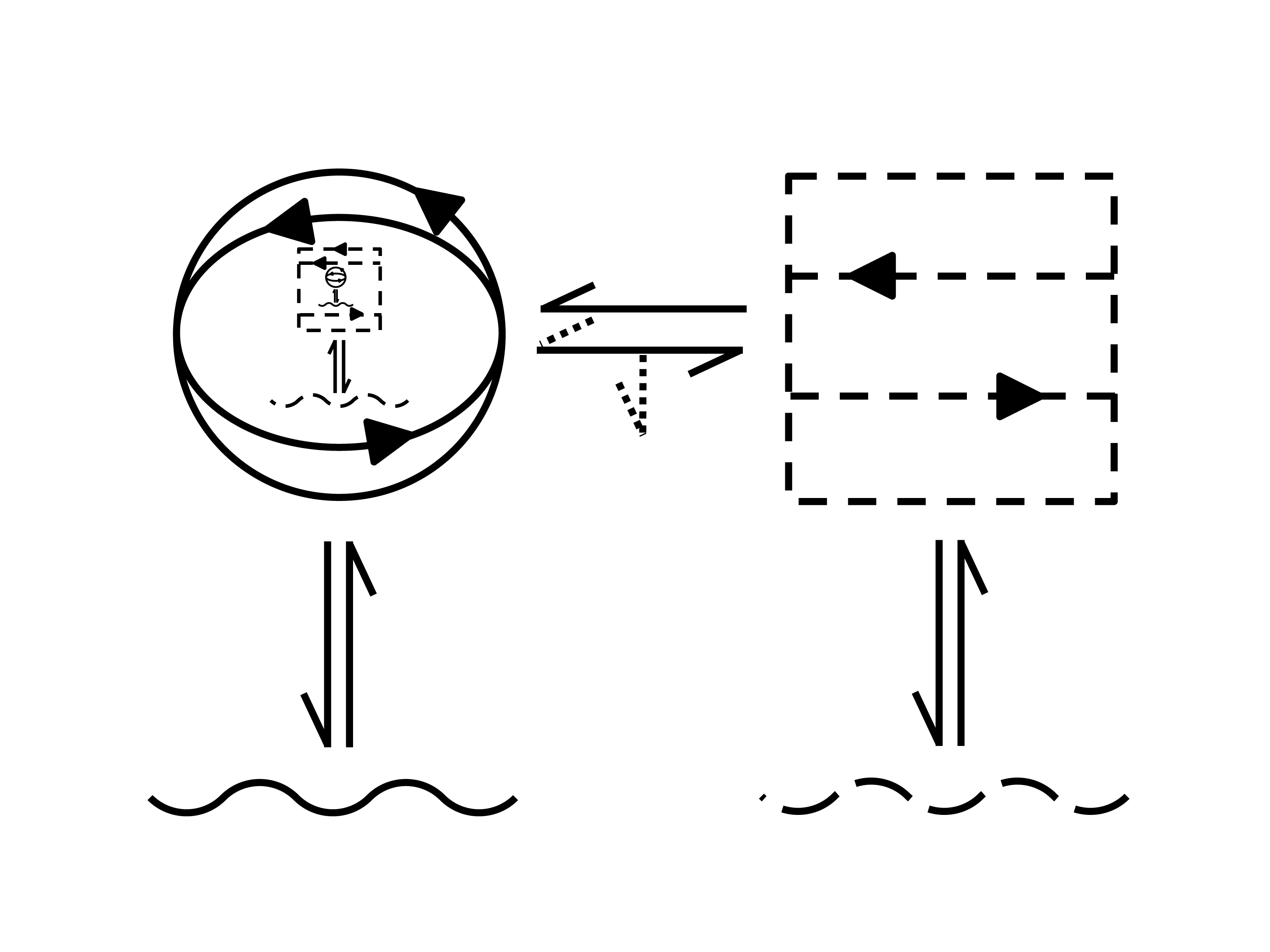}
\caption{Pictographic representation of coupling between a human being (on the left) and a cognitive robot (on the right).  The robot lacks the capability of ostensive recursive mind-reading (it has no Theory-of-Mind), so the interaction is inevitably constrained.}
\label{fig:MIS}
\end{figure}

So now we arrive at an interesting position; if (spoken) language interaction between human beings is grounded through shared experiences, representations and priors, to what extent is it possible to construct a technology that is intended to replace one of the participants?  For example, if one of the interlocutors illustrated in Fig.\ref{fig:DIA} is replaced by a cognitive robot (as in Fig.\ref{fig:MIS}), then there will be an inevitable mismatch between the capabilities of the two partners, and coupled ostensive recursive mind-reading (i.e. full language) cannot emerge.

Could it be that there is a fundamental limit to the language-based interaction that can take place between \emph{unequal} partners - between humans and machines?  Indeed, returning to the question posed in Section \ref{sec:HLF} ``\emph{Is there such a thing as half a language?}'', the answer seems to be ``\emph{no}''; spoken language does appear to be all-or-nothing \ldots

\begin{quotation}
``\emph{The assumption of continuity between a fully coded communication system at one end, and language at the other, is simply not justified.}'' \cite{ScottPhillips2015} (p.46).
\end{quotation}

\section{The Way Forward?} \label{sec:WF}

The story thus far provides a compelling explanation of the less-than-satisfactory experiences enjoyed by existing users of speech-enabled systems and identifies the source of the \emph{habitability gap} outlined in Section \ref{sec:HG}.  It would appear that, due to the gross mismatch between their respective priors, it might be impossible to create an automated system that would be capable of a sustained and productive language-based interaction with a human being (except in narrow specialised domains involving experienced users).  The vision of constructing a general-purpose voice-enabled autonomous social agent may be fundamentally flawed - the equivalent of trying to build a vehicle that travels faster than light!

However, before we give up all hope, it is important to note that there are situations where voice-based interaction between mismatched partners \emph{is} successful - but these are very different from the scenarios that are usually considered when designing current speech-based systems.  For example, human beings regularly engage in vocal interaction with members of a different cultural and/or linguistic and/or generational background\footnote{Interestingly, Nass \& Brave \cite{Nass2005} noted that people speak to poor automatic speech recognition systems as if they were non-native listeners.}.  In such cases, all participants dynamically adjust many aspects of their behaviour - the clarity of their pronunciation, their choice of words and syntax, their style of delivery, etc. - all of which may be controlled by the perceived effectiveness of the interaction (that is, using \emph{feedback} in a coupled system).  Indeed, a particularly good example of such accommodation between mismatched interlocutors is the different way in which caregivers talk to young children (termed ``\emph{parentese}'') \cite{Fernald1985}.  Maybe these same principles should be applied to speech-based human-machine interaction?  Indeed, perhaps we should be explicitly studying the particular adaptations that human beings make when attempting to converse with autonomous social agents - a new variety of spoken language that could be appropriately termed ``\emph{robotese}''\footnote{Unfortunately, this term has already been coined to refer to a robot's natural language abilities in robot-robot and robot-human communication \cite{Matson2011}.}.

Of course, these scenarios all involve spoken interaction between one human being and another, hence in reality there is a huge overlap of priors in terms of bodily morphology, environmental context and cognitive structure, as well as learnt social and cultural norms.  Arguably the largest mismatch arises between an adult and a very young child, yet this is still interaction between members of the same species.  A more extreme mismatch exists between non-conspecifics; for example, between humans and animals.  However, it is interesting to note that our nearest relatives - the apes - do not have language, and this seems to be because they do not have the key precursor to language: ostensive communication (apes do not seem to understand pointing gestures) \cite{ScottPhillips2015}.

Interestingly, one animal - the domestic dog - appears to excel in ostensive communication and, as a consequence, dogs are able to engage in very productive spoken language interaction with human partners (albeit one-sided and somewhat limited in scope) \cite{Serpell1995,ScottPhillips2015}.  Spoken human-dog interaction may thus be a potentially important example of a heavily mismatched yet highly effective cooperative configuration that might usefully inform spoken human-robot interaction in hitherto unanticipated ways.

\section{Final Remarks} \label{sec:CONC}

This paper has argued that there is a fundamental \emph{habitability} problem facing contemporary spoken language systems, particularly as they penetrate the mass market and attempt to provide a general-purpose voice-based interface between human users and (so-called) intelligent systems.  It has been suggested that the source of the difficulty in configuring genuinely usable systems is twofold: first, the need to align the visual, vocal and behavioural affordances of the system, and second, the need to overcome the huge mismatch between the capabilities and expectations of a human being and the features and benefits offered by even the most advanced autonomous social agent.  This led to the preliminary conclusion that spoken language may indeed be all-or-nothing.

Finally, and on a positive note, it was observed that there are situations where successful spoken language interaction can take place between mismatched interlocutors (such as between native and non-native speakers, or between an adult and a child, or even between a human being and a dog).  It is thus concluded that these scenarios might provide critical inspiration for the design of future speech-based human-machine interaction.

\subsection*{Acknowledgement}

This work was supported by the European Commission [grant numbers EU-FP6-507422, EU-FP6-034434, EU-FP7-231868 and EU-FP7-611971], and the UK Engineering and Physical Sciences Research Council [grant number EP/I013512/1].

\end{document}